\documentclass[prl,aps,amsmath,amsfonts,showpacs,twocolumn,floatfix,superscriptaddress]{revtex4-1}

\usepackage{color}
\usepackage{bm}
\usepackage{graphicx}

\def\be{\begin{equation}}
\def\ee{\end{equation}}
\newcommand{\corr}[1]{\langle #1\rangle}

\def\Re{\mathop{\rm Re}}
\def\Im{\mathop{\rm Im}}
\def\eps{\varepsilon}
\def\br{{\bf r}}

\newcommand{\lee}{l_\text{\emph{e-e}}}
\newcommand{\lep}{l_\text{\emph{e}-ph}}

\newcommand{\tep}{$e$-ph}
\newcommand{\ETh}{E_\text{Th}}

\def\Vinst{V_\text{inst}}

\def\Ai{\mathop{\rm Ai}}

\begin{document}

\title{Onset of superconductivity in a voltage-biased NSN microbridge}

\author{Maksym Serbyn}
\affiliation{Department of Physics, Massachusetts Institute of
Technology, Cambridge, Massachusetts 02139, USA}
\affiliation{L. D. Landau Institute for Theoretical Physics, Chernogolovka,
Moscow Region, 142432, Russia}
\author{Mikhail A. Skvortsov}
\affiliation{L. D. Landau Institute for Theoretical Physics, Chernogolovka,
Moscow Region, 142432, Russia}
\affiliation{Moscow Institute of Physics and Technology, Moscow 141700, Russia}

\date{\today}

\begin{abstract}
We study the stability of the normal state in a mesoscopic NSN
junction biased by a constant voltage $V$ with respect to the
formation of the superconducting order. Using the linearized
time-dependent Ginzburg-Landau equation, we obtain the temperature
dependence of the instability line, $\Vinst(T)$, where nucleation of
superconductivity takes place. For sufficiently low biases, a
stationary symmetric superconducting state emerges below the
instability line. For higher biases, the normal phase is destroyed
by the formation of a non-stationary bimodal state with two
superconducting nuclei localized near the opposite terminals.
The low-temperature and large-voltage behavior of the instability line
is highly sensitive to the details of the inelastic relaxation mechanism in the wire.
Therefore, experimental studies of $\Vinst(T)$
in NSN junctions may be used as an effective tool to access
parameters of the inelastic relaxation in the normal state.
\end{abstract}

\pacs{
74.40.Gh, % nonequilibrium processes in superconductivity,
74.78.Na,  %superconducting mesoscopic systems
72.15.Lh, %Relaxation processes in electrical conductivity (metals and alloys)
72.10.Di % e-ph interactions in electronic transport
}

\maketitle

Nonequilibrium superconductivity has being attracting significant experimental and theoretical attention over decades~\cite{Kopnin-book,Kapitza,GulianZharkov},
ranging from vortex dynamics~\cite{Larkin}
to the physics of the resistive state in current-carrying superconductors~\cite{VodolazovPRL03,XiongPRL97,RogachevPRL06,PengPRB11,AstafievNature12}.
It was recognized long ago~\cite{IvlevKopnin}
that a superconducting wire typically has a hysteretic current voltage
characteristic specified by several ``critical'' currents.
In an up-sweep, a current exceeding the thermodynamic
depairing current, $I_c(T)$, does not completely destroy superconductivity
but drives the wire into a nonstationary resistive state~\cite{Meyer},
with the excess phase winding relaxing through the formation
of phase slips~\cite{LangerAmbegaokar, *McCumberHalperin}. The resistive state continues until $I_2(T)>I_c(T)$,
when the wire eventually becomes normal.
In the down-sweep of the current voltage
characteristic, the wire remains normal until $I_1(T)<I_2(T)$
when an emerging order parameter leads to the reduction
of the wire resistance.

The theoretical description of a nonequilibrium superconducting state
is a sophisticated problem, requiring a simultaneous
account of the nonlinear order parameter dynamics and quasiparticle relaxation
under nonstationary conditions. The resulting set of equations is extremely
complicated \cite{Larkin,Kopnin-book} and can be treated only numerically
\cite{KeizerPRL06,Klapwijk12,SnymanPRB09}
(even then the stationarity of the superconducting state
is often assumed~\cite{KeizerPRL06,Klapwijk12}).
A more intuitive but somewhat oversimplified approach
is based on the a time-dependent Ginzburg-Landau (TDGL) equation
for the order parameter field $\Delta(\br,t)$.
The TDGL approach which is generally inapplicable in the gapped phase~\cite{EliashbergTDGL},
can be justified only in a very narrow vicinity of the critical temperature, $T_c$,
provided that the electron-phonon (\tep) interaction
is sufficiently strong to thermalize quasiparticles~\cite{KramerPRL78}.
These generalized TDGL equations are analyzed numerically in
Refs.~\cite{VodolazovPRL03,VodolazovPRB04, *VodolazovPRB07-2, *VodolazovPRB08}.

While the applicability of the TDGL equation in the superconducting region
is a controversal issue, its linearized form can be safely employed
to find the line $I_\text{inst}(T)$ of the absolute instability
of the normal state with respect to the appearance of an infinitesimally small
order parameter
$\Delta(\br,t)$~\cite{IvlevKopnin,RubinsteinPRL07,KolyaEPL09}.
If the transition to the superconducting state is second order, then
$I_1(T)$ coincides with $I_\text{inst}(T)$.  Otherwise the actual
instability takes place at a larger $I_1(T)>I_\text{inst}(T)$. In
both cases, $I_\text{inst}(T)$ gives the lower bound for $I_1(T)$.

Previous results~\cite{IvlevKopnin,RubinsteinPRL07}
for the instability line of a superconducting wire
connected to normal reservoirs (NSN microbridge)
have been obtained in the limit of quasi-equilibrium,
when strong \tep{} relaxation renders the distribution function
locally thermal.
This approximation breaks down for low-$T_c$
superconducting wires shorter than the {\tep} relaxation length,
$\lep(T_c)$ (e.g., for aluminium, $\lep(T_c)\approx40$ $\mu$m
\cite{Giazotto-RMP}). Such systems have recently been
experimentally studied in Refs.~\cite{Klapwijk12} (Al) and
\cite{YuChenPRL09,TianPRL05} (Zn; reservoirs may be driven normal
by a magnetic field). It was found that for sufficiently large
biases superconductivity arises near the terminals through a
second-order phase transition, with $I_1(T)=I_\text{inst}(T)$~\cite{Klapwijk12}.

In this Letter we study the normal state instability line in
an
NSN microbridge biased by a DC voltage $V$,
relaxing the assumption of strong thermalization.
For small biases, $eV\ll T_c$, the instability line is universal and we reproduce
the results of
Refs.~\cite{IvlevKopnin,RubinsteinPRL07}.
The universality breaks down for larger biases, where we obtain $V_\text{inst}(T)$
\emph{as a functional of the normal state
distribution function}\/ and analyze it for various types of
inelastic interactions.

We model the NSN microbridge as a diffusive wire of length $L$
coupled at $x=\pm L/2$ to large normal reservoirs via transparent
interfaces. The terminals are biased by a constant voltage $V$.
The wire length, $L$, is assumed to be larger than
the zero-temperature coherence length, $\xi_0=\sqrt{\pi D/8T_{c0}}$,
where $D$ is diffusion coefficient,
and $T_{c0}$ is the critical temperature of the infinite wire.
The equilibrium critical temperature,
$T_c = T_{c0}(1-\pi^2\xi_0^2/L^2)$, is smaller than $T_{c0}$
due to the finite size effect~\cite{Boogaard}.
In what follows we neglect superconducting fluctuations~\cite{LV}.

\emph{General stability criterion.}---%
An arbitrary nonequilibrium normal state becomes
absolutely unstable with respect to superconducting fluctuations
if an infinitesimally small order parameter, $\Delta(\br,t)$, does
not decay with time but evolves to finite values. Evolution of
$\Delta(\br,t)$ is described by the TDGL equation, and it
suffices to keep only the linear term to judge its stability.
In dirty superconductors, the linearized TDGL equation can be readily
derived from the Keldysh $\sigma$-model formalism~\cite{LevchenkoPRB07,FLS00,Kamenev}
or dynamic Usadel equations~\cite{Larkin}
by expanding in $\Delta$.
It takes the form $(L_R)^{-1}*\Delta=0$, where $(L^R)^{-1}$
is the inverse fluctuation propagator, and convolution in time
and space indices is implied. In the frequency representation,
$(L^R_\omega)^{-1}$ is an integral operator in real space specified
by the kernel
\begin{equation} \label{Eq:LRdef}
  (L^{R}_\omega)^{-1}_{\br,\br'}
  =
  -\frac{\delta_{\br,\br'}}{\lambda}
  +
  i
  \int_{-\omega_D}^{\omega_D}
  dE \,
  F(E,\br) \,
  C_{\omega-2E}(\br,\br') ,
\end{equation}
where $\lambda$ is the dimensionless BCS interaction constant,
$\omega_D$ is the Debye frequency, and $C$ stands for the retarded Cooperon,
$C_\eps(\br,\br') = \corr{\br|(-D\nabla^2-i\eps)^{-1}|\br'}$,
vanishing at the boundary with the terminals.

The operator (\ref{Eq:LRdef}) is a functional of the normal-state
nonequilibrium electron distribution function, $F(E,\br)$, which contains
information about the mechanism for inelastic relaxation in the wire.
The distribution function should be determined from the kinetic equation
\be
\label{kinur}
  D\nabla^2F(E,\br)+{\cal I}^\text{\emph{e-e}}[F]+{\cal I}^\text{\emph{e}-ph}[F]=0,
\ee
with ${\cal I}^\text{\emph{e-e}}[F]$ and ${\cal I}^\text{\emph{e}-ph}[F]$
being the electron-electron \mbox{(\emph{e-e})}
and \emph{e}-ph  collision integrals, respectively.
The corresponding energy relaxation lengths,
$\lee(T)\propto T^{-1/4}$ and $\lep(T)\propto T^{-3/2}$,
behave as a negative power of the temperature $T$ in quasi-equilibrium~\cite{Giazotto-RMP}.

In the absence of inelastic collisions,
the kinetic equation (\ref{kinur}) is solved by
the ``two-step'' function \cite{Nagaev95,PothierPRL97}:
\begin{equation} \label{Eq:Ftwostep}
  F(E,x)
  =
  (1/2-x/L) F_L(E)
  +
  (1/2+x/L) F_R(E)
  .
\end{equation}
The distribution functions in the terminals,
$F_{L,R}(E)=F_0(E\pm eV/2)$, are given by the equilibrium distribution
function, $F_0(E)=\tanh(E/2T)$, shifted by $\pm eV/2$~($e>0$).
In the opposite case of strong inelastic relaxation,
the distribution function takes the form
\begin{equation}\label{Eq:Fee-ep}
  F_\text{in}(E,x)
  =
  \tanh[(E-e\phi(x))/2T(x)]
  ,
\end{equation}
where $\phi(x)=Vx/L$ is the potential in the normal state,
and $T(x)$ is the effective temperature. For strong lattice
thermalization ($\lep \ll L \ll \lee$), $T(x)=T$.
For the dominating \emph{e-e} scattering ($\lee\ll L \ll\lep$), $T^2(x) = T^2 + (3/4\pi^2) [1-(2x/L)^2] (eV)^2$~\cite{Nagaev95}.

The order parameter evolution governed by the linearized TDGL
operator~(\ref{Eq:LRdef}) can be naturally described in terms of the
eigenmodes $\Delta_k(\br)e^{-i\omega_k t}$ annihilated by
$(L^R_\omega)^{-1}$. The normal state is stable provided
$\Im\omega_k<0$ for all eigenmodes.
Below we analyze the spectrum of the operator (\ref{Eq:LRdef})
and determine the instability line in the coordinates $T$--$V$.
In the general case, the spectrum can be obtained only numerically.
Analytical considerations become possible when the operator~(\ref{Eq:LRdef})
may be linearized in $\omega$: $(L^R_\omega)^{-1} = i\tau\omega - {\cal H}$.
The instability occurs when the real part of the lowest eigenvalue
of the operator $\cal H$ turns to zero.

\emph{Weak-nonequilibrium regime.}---%
In the limit of low biases, $eV\ll T_c$, the deviation from
equilibrium is small everywhere in the wire and
the distribution function acquires a universal form,
$F(E,x) \approx F_0(E) - F_0'(E) \, e\phi(x)$,
regardless of the relaxation mechanism.
Then Eq.~(\ref{Eq:LRdef}) takes the form
$
  (L^R_\omega)^{-1}
  =
  i\pi \omega/8T
  -\ln(T/T_{c0})
  -
  (\xi_0/L)^2 \,
  {\cal H}_v
$,
with
\be
\label{Eq:H-iv}
  {\cal H}_v
  =
  - \partial_{\tilde x}^2
  + 2 i v \tilde x
  ,
\qquad
  \tilde x \in [-1/2,1/2] .
\ee
The Hamiltonian ${\cal H}_v$ describes quantum-mechanical motion
\emph{in an imaginary electric field}, $v=eV/\ETh$,
on the interval $\tilde x \equiv x/L \in [-1/2,1/2]$,
where $\ETh = D/L^2$ is the Thouless energy.
Hard-wall boundary conditions, $\psi(\pm1/2) = 0$, imposed on wave
functions correspond to the complete suppression of superconductivity
at the contacts with highly-conducting terminals due to the inverse
proximity effect.
\begin{figure}
\includegraphics[width=\linewidth]{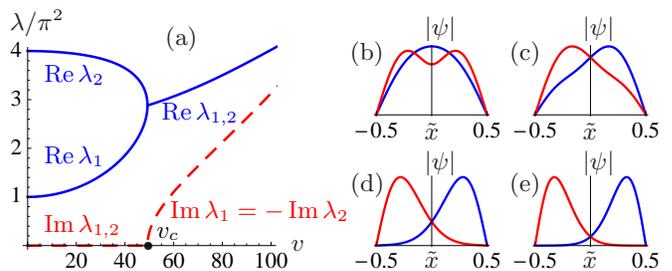}
\caption{(Color online)
(a) Real (solid blue line) and imaginary (dashed red line) parts of the lowest
eigenvalues $\lambda_{1,2}(v)$ of the Hamiltonian (\ref{Eq:H-iv}).
The spectrum is entirely real until $v=v_c\approx49.25$.
(b)--(e) Spatial dependence of the absolute values of the eigenfunctions, $|\psi_{1}({\tilde x})|$~(blue) and $|\psi_{2}({\tilde x})|$~(red),
for $v=0.8\,v_c$, $1.2\,v_c$, $5\,v_c$, and $10\,v_c$ respectively.
} \label{F:functions}
\end{figure}

The Hamiltonian (\ref{Eq:H-iv}) has been recently
analyzed in Ref.~\cite{RubinsteinPRL07}. It belongs to a class of
non-Hermitian Hamiltonians invariant under the combined action of
the time-reversal, ${\cal T}$: $f(x)\mapsto f^*(x)$, and parity,
${\cal P}$: $f(x)\mapsto f(-x)$, transformations. The $\cal
PT$-symmetry of the Hamiltonian~(\ref{Eq:H-iv}) ensures that its
eigenvalues $\lambda_n(v)$ are either real or form
complex-conjugated pairs, providing a complex extension
of the notion of Hermiticity~\cite{BenderPRL98,BenderRPP07}.
At $v=0$, the spectrum is non-degenerate: $\lambda_n(0)=\pi^2n^2$
($n=1,2,\dots$). It evolves
continuously with $v$ and a nonzero $\Im\lambda(v)$ arises only when
the two lowest eigenvalues, $\lambda_1(v)$ and $\lambda_2(v)$,
coalesce~[see Fig.~\ref{F:functions}~(a)]. This happens at
$v=v_c\approx 49.25$~\cite{RubinsteinPRL07}, indicating the
transition to a complex-valued spectrum. For $v<v_c$, the ground
state of (\ref{Eq:H-iv}) is $\cal PT$-symmetric, and hence
$|\psi_1(\tilde x)|=|\psi_1(-\tilde x)|$.
For $v>v_c$, the $\cal PT$-symmetry is spontaneously broken and
there is a pair of states with the lowest $\Re\lambda(v)$:
$\psi_L(\tilde x)=\psi_1(\tilde x)$ and $\psi_R(\tilde
x)=\psi_2(\tilde x)=\psi_1^*(-\tilde x)$, shifted to the left
(right) from the midpoint [see Fig.~\ref{F:functions} (b)--(e)].

\begin{figure}
\includegraphics[width=\linewidth]{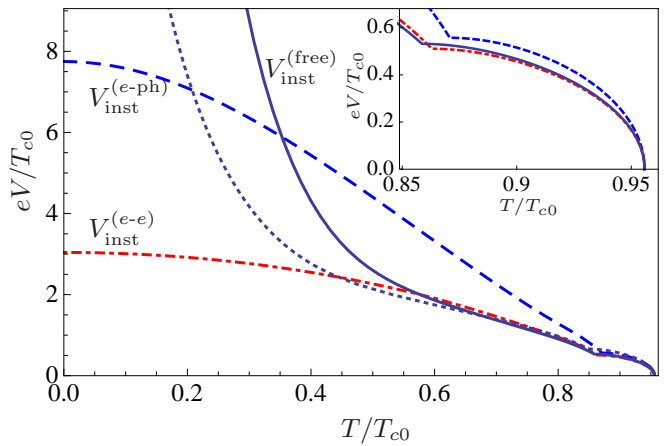}
\caption{(Color online) Instability
voltage as a function of temperature, $\Vinst(T)$, obtained numerically for a wire of length $L=15 \xi_0$ for
three limiting types of the distribution function: without inelastic
relaxation (solid blue line), and with dominant $e$-$e$
(dot-dashed line) or $e$-ph~(dashed line) relaxation.
Dotted curve illustrates the suppression of $\Vinst^\text{(free)}(T)$
by a finite terminal resistance, $\beta=0.1$ (see text).
Inset shows the behavior in the vicinity of the bifurcation point
$eV_*\approx 50 \ETh \approx 0.57\, T_{c0}$.
}
\label{F:Vc}
\end{figure}

Spontaneous breaking of the $\cal PT$-symmetry associated with
the spectral bifurcation at $v=v_c$ explains the appearance of
asymmetric superconducting states observed in numerical
simulations~\cite{VodolazovPRB07} and recent
experiments~\cite{Klapwijk12}. The normal-state instability line,
$\Vinst(T)$, is specified implicitly by the relation
\begin{equation}\label{Eq:Tc(V)small}
  1
  -
  T/T_{c0}
  =
  (\xi_0/L)^2 \Re \lambda_1(e\Vinst(T)/\ETh),
\end{equation}
and exhibits a singular behavior at the critical bias $V_*$ given by
$eV_*=v_c\ETh \approx 50\ETh$ (see inset in Fig.~\ref{F:Vc}).
The bifurcation of the instability line occurs at the temperature
$T_*\approx T_{c0}(1- 28.44\,\xi_0^2/L^2)$.
For long wires ($L\gg\xi_0$), $T_*$ is very close to $T_c$.

The time dependence of the emergent superconducting state
is determined by $\Im\lambda_1(v)$.
Below the bifurcation threshold, for $\Vinst(T)<V_*$,
the system undergoes at $V=\Vinst(T)$ the transition to a \emph{stationary}\/
superconducting state, with the superconducting chemical potential
being the half-sum of the chemical potentials in the terminals.
This state is supercurrent-carrying,
and can withstand a maximum phase winding of $\pi$ achieved at
the critical bias $V_*$. For larger voltages, $\Vinst(T)>V_*$, two
modes, $\psi_L(x)$ and $\psi_R(x)$, nucleate simultaneously at
$\Vinst(T)$. The resulting bimodal superconducting state is
\emph{non-stationary} since the left and right modes feel
different electrochemical potentials and their phases rotate with
opposite frequencies, $\Omega_{L,R}(V)=\mp\ETh\Im\lambda_1(eV/T)$.
This will result in the Josephson generation with the differential
frequency $\Omega(V)=\Omega_R(V)-\Omega_L(V) \propto\sqrt{V-V_*}$
as $\Vinst(T)\to V_*$.
Though these supercurrent oscillations are locked to the superconducting part
they may excite oscillations of the normal current in the whole circuit.
Thus the dc biased NSN microbridge may act as a voltage-tunable generator
of an ac current, with the maximal amplitude of
oscillations
expected in the \emph{coherent} regime, $V-V_*\sim V_*$.
The possibility of experimental observation of such a generation
remains an open problem.

\emph{Incoherent regime.}---%
As the voltage is increased far above the bifurcation threshold, $\Vinst(T)\gg V_*$,
the eigenmodes $\psi_{L,R}(x)$ gradually localize
near
the corresponding terminals, with their size, $a(V)$,
becoming much smaller then the wire length [see Fig.~\ref{F:functions}~(b)--(e)].
This is the \emph{incoherent} regime, where the overlap between
$\psi_L(x)$ and $\psi_R(x)$ is exponentially small, supercurrent
oscillations are suppressed, and nucleation of superconductivity
near each terminal can be described independently~\cite{Klapwijk12}.

Using $a(V)/L$ as a small parameter and still working in the
vicinity of $T_c$, we linearize $F(E,x)$ near the left
terminal and reduce Eq.~(\ref{Eq:LRdef}) to the form:
$
  (L^R_\omega)^{-1}
  =
  i\pi (\omega+eV)/8T
  -\ln(T/T_{c0})
  - {\cal H}_\alpha
$,
where the operator
\be
\label{Ha}
  {\cal H}_\alpha
  =
  - \xi_0^2 \partial_{x_L}^2
  + \alpha x_L ,
\qquad
  x_L \geq 0 ,
\ee
acts on the semiaxis $x_L\equiv x+L/2\geq0$ with the boundary
condition $\psi(0)=0$. The complex parameter $\alpha$
is a functional of the distribution function:
\begin{equation}\label{Eq:alphadef}
  \alpha\Bigl(\frac{eV}{T}\Bigr)
  =
  -
  \int
  \frac{dE \, \partial_x F(E-eV/2,x)|_{x=-L/2+a(V)}}
  {2(E-i0)}
  .
\end{equation}

The ground state of the Hamiltonian (\ref{Ha}) has the energy
$\gamma_0(\alpha\xi_0)^{2/3}$ and the wave function
$\psi_0(x_L)=\Ai(x_L/a-\gamma_0)/\sqrt{a}\Ai'(-\gamma_0)$,
where $-\gamma_0\approx-2.34$ is the first zero of the Airy function,
and $a=(\xi_0^2/\alpha)^{1/3}$ is the nucleus size~\cite{IvlevKopnin}.
For the instability line we get:
\be
\label{Eq:Tc(V)large}
  1 -
  T/T_{c0}
  =
  \gamma_0
  \xi_0^{2/3}
  \Re \alpha^{2/3}(e\Vinst(T)/T)
  .
\ee
The left and right unstable states rotate with the frequencies
$\Omega_{L,R}(V) = \mp[eV-\Omega_1(V)]$,
where
$
  \Omega_1(V)
  =
  (8T_{c0}/\pi)
  \gamma_0
  \xi_0^{2/3}
  \Im \alpha^{2/3}(eV/T)
$
is a small correction to the Josephson frequency determined
by the electrochemical potential of the corresponding terminal.
At the instability line, $V=\Vinst(T)$, the size $a$ of the unstable mode
is of the order of the temperature-dependent superconducting
coherence length $\xi(T)\sim (1-T/T_{c0})^{-1/2}\xi_0$.

For long wires ($L\gg\xi_0$),
the incoherent regime partly
overlaps with the weak-nonequilibrium regime.
Then for $eV_*\ll e\Vinst(T)\ll T_c$,
Eq.~(\ref{Eq:Tc(V)large}) gives a universal answer
\begin{equation}\label{Eq:VcsmallV}
  \frac{e\Vinst(T)}{T_{c0}}
  =
  \frac{2^{7/2}}{\pi} \frac{L}{\xi_0}
  \left( \frac{T_{c0}-T}{\gamma_0 T_{c0} }\right)^{3/2}
  ,
\end{equation}
which could have also been deduced from Eq.~(\ref{Eq:Tc(V)small})
using the quasiclassical approximation for $\lambda_1(v)$ at $v\gg1$.
Equation (\ref{Eq:Tc(V)large}) exactly coincides with the result
of Ref.~\cite{IvlevKopnin}, predicting
that superconductivity nucleates near the terminals at a finite
current $I_\text{inst}(T)\approx 0.356\, I_c(T)$.

The position of the instability line in the incoherent regime at large biases,
$e\Vinst(T)\gg T_c$, depends on the relation between the inelastic lengths
$\lee$ and $\lep$, the wire length $L$, \emph{and}\/ the nucleus size $a(V)$.
The presence of the latter scale, which probes the distribution
function near the boundaries of the wire, leads to a rich variety of regimes
realized at different temperatures.

For the three limiting distributions [Eqs.~(\ref{Eq:Ftwostep})
and (\ref{Eq:Fee-ep})], the function $\alpha(u)$ can be found analytically:
(i) $\alpha_\text{free}(u) = [\psi(1/2+iu/2\pi)-\psi(1/2)]/L$
for the non-interacting case, $L\ll\lee,\lep$, where $\psi(x)$ is the digamma function;
(ii) $\alpha_{\text{\emph{e}-ph}}(u) = i \pi u/4L$
for strong lattice thermalization, $\lep\ll a(V)\ll L\ll\lee$;
and (iii) $\alpha_\text{\emph{e-e}}(u) = [i \pi u/4 + 3 u^2/2\pi^2]/L$
for the dominant $e$-$e$ interaction, $\lee\ll a(V)\ll L\ll\lep$.
In case (ii), the instability line $\Vinst^\text{$e$-ph}(T)$ is
given by Eq.~(\ref{Eq:VcsmallV}).
In the vicinity of $T_c$, the instability lines in cases (i) and (iii) are given by:
\begin{gather}
\label{Eq:Vc-elastic}
  \frac{e\Vinst^\text{(free)}(T)}{T_{c0}}
  =
  1.13
  \exp\left\{
    \frac{L}{\xi_0}
    \left(\frac{T_{c0}-T}{\gamma_0 T_{c0}}\right)^{3/2}
  \right\}
  ,
\\
\label{Eq:Vc-ee}
  \frac{e\Vinst^\text{($e$-$e$)}(T)}{T_{c0}}
  =
  \left( \frac{2\pi^2}{3} \frac{L}{\xi_0} \right)^{1/2}
  \left( \frac{T_{c0}-T}{\gamma_0 T_{c0} }\right)^{3/4}
  .
\end{gather}
Counterintuitively, in cases (i) and (iii) the instability current
$I_\text{inst}(T)\propto\Vinst(T)/L$ has a nontrivial dependence
on the system size, as opposed to Eq.~(\ref{Eq:VcsmallV}).
Such a behavior is a consequence of strong nonequilibrium in the wire.
The limiting curves $\Vinst^\text{(free)}(T)$,
$\Vinst^\text{($e$-ph)}(T)$, and $\Vinst^\text{($e$-$e$)}(T)$
for all temperatures obtained numerically from Eq.~(\ref{Eq:LRdef})
for the wire with $L/\xi_0=15$ are shown in Fig.~\ref{F:Vc}.
The universal behavior at small biases can be easily seen
(inset).
Since the ratio $L/\xi_0$ is not very large, the instability line
becomes strongly dependent on the distribution function
already for $V\gtrsim V_*$.

The most exciting feature of our results is the exponential
growth of $\Vinst(T)$ with decreasing temperature
in the non-interacting case, Eq.~(\ref{Eq:Vc-elastic}).
Hence, even a small deviation of the distribution function
from the two-step form~(\ref{Eq:Ftwostep}) will drastically
modify $\Vinst(T)$.
As an example, consider the effect of a finite resistance
of the normal terminals.
Then the function $F_L(E)$
in Eq.~(\ref{Eq:Ftwostep})
will be replaced by
$F_{L}(E) = \beta F_0(E+ eV/2)+ (1-\beta)F_0(E- eV/2)$,
where $V$ is the voltage applied to the NSN microbridge,
and $\beta=R_T/(R_N+2R_T)$~[$R_T$ and $R_N$ are
the resistances of the N and S part of the junction, respectively].
The resulting $\Vinst(T)$ for $\beta=0.1$
is shown by the dotted blue line in Fig.~\ref{F:Vc}.
While $\Vinst(T)$ is unchanged for small biases,
it is strongly suppressed compared to
$\Vinst^\text{(free)}(T)$ for large biases.

\emph{Low-temperature behavior.}---%
The exponential growth of $\Vinst^\text{(free)}(T)$ in the non-interacting
case formally implies that superconductivity at
$T=0$
might persist up to exponentially large voltages,
$\ln(e\Vinst(0)/T_{c0}) \sim L/\xi_0 \gg 1$.
This conclusion is wrong, since
inelastic relaxation and heating
become important with increasing $V$,
even if they were negligible at $V=0$.
To study the low-$T$ part of the instability line,
we consider here a model of the \emph{e}-ph interaction
(\emph{e-e} relaxation neglected) when the phonon temperature
is assumed to coincide with the base temperature of the terminals
and \emph{e}-ph relaxation is weak at $T_c$: $\lep(T_c)\gg L$
(as in the experiment~\cite{Klapwijk12}).

With decreasing $T$ below $T_c$, the instability line first
follows Eq.~(\ref{Eq:Vc-elastic}). At the same time,
$\lep$ decreases and eventually the distribution function in the
middle of the wire becomes nearly thermal with the effective
temperature $T_\text{eff}$. This happens when $T_\text{eff}$
obtained from the heat balance equation~\cite{Giazotto-RMP},
$(eV/L)^2\sim T_\text{eff}^5/T_c^3\lep^2(T_c)$, becomes so large
that $\lep(T_\text{eff})\sim L$.
The corresponding voltage, $V_\text{ph}$, can be estimated as
$eV_\text{ph}/T_c\sim[\lep(T_c)/L]^{2/3}$.
Consequently, the exponential growth~(\ref{Eq:Vc-elastic})
persists for voltages
$V_* \lesssim V \lesssim V_\text{ph}$, corresponding to
the temperature range $T_\text{ph}\lesssim T \lesssim T_*$,
where with logarithmic accuracy $1-T_\text{ph}/T_{c0} \sim (\xi_0/L)^{2/3}$.

For higher biases, $V>V_\text{ph}$,
electrons in the central part of the wire
have the temperature $T_\text{eff}$. However, the parameter~$\alpha$,
Eq.~(\ref{Eq:alphadef}), is determined by the distribution
function in the vicinity of the terminals which is not
thermal. Matching solution of the collisionless kinetic equation
for $0<x_L<\lep(T_\text{eff})$ at the effective right
``boundary'', $x_L=\lep(T_\text{eff})$, with the function
(\ref{Eq:Fee-ep}) with $T(x)=T_\text{eff}$, we obtain
$\alpha\sim1/\lep(T_\text{eff})$. Therefore, for $V\gtrsim
V_\text{ph}$ we get with logarithmic accuracy:
\be
\label{Vc-low-T}
  \frac{e\Vinst(T)}{T_{c0}}
  \sim
  \frac{L}{\xi_0}
  \left(\frac{\lep(T_c)}{\xi_0}\right)^{2/3}
  \left(\frac{T_{c0}-T}{T_{c0}}\right)^{5/2} .
\ee
Equation (\ref{Vc-low-T}) corresponding to the case $a(V)\ll \lep\ll L$
is different from the expression (\ref{Eq:VcsmallV}) when phonons
are important already at $T_c$, and $\lep\ll a(V)\ll L$.
The scaling dependence of Eq.~(\ref{Vc-low-T}) on $L$ indicates that the stability
of the normal state is controlled by the applied current,
similar to Eq.~(\ref{Eq:VcsmallV}).
At zero temperature the instability current exceeds the thermodynamic depairing current
by the factor of $(\lep(T_c)/\xi_0)^{2/3}\gg1$.

\emph{Discussion.}---%
Our general procedure locates the absolute instability line, $\Vinst(T)$,
of the normal state for a voltage-biased NSN microbridge.
Following experimental data~\cite{Klapwijk12}
we assumed that the onset of superconductivity is of the second order.
While non-linear terms in the TDGL equation are required
to determine the order of the
phase
transition~\cite{tobe},
we note that were it of the first order,
its position would be shifted to voltages higher than~$\Vinst(T)$.

In the vicinity of $T_c$, the problem of finding $\Vinst(T)$ can
be mapped onto a one-dimensional quantum mechanics in some
potential $U(x)$.
For small biases, $eV \ll T_{c0}$,
the potential $U(x)$ does not depend on the
distribution function details, explaining universality of the
instability line, including the bifurcation from the single-mode
to the bimodal superconducting state at $eV\sim 50
\ETh$~\cite{RubinsteinPRL07} and nucleation of superconductivity
in the vicinity of the terminals for larger biases \cite{IvlevKopnin}.

For $eV \gtrsim T_{c0}$, the potential $U(x)$ becomes
a functional of the normal-state distribution function, producing
$\Vinst(T)$ that is strongly sensitive to inelastic relaxation
mechanisms in the wire. For the dominant \tep{} interaction, the
instability is controlled by the electric field ${\cal E} = V/L$
[Eqs.~(\ref{Eq:VcsmallV}) and (\ref{Vc-low-T})], while in the
opposite case [Eqs.~(\ref{Eq:Vc-elastic}) and (\ref{Eq:Vc-ee})],
the instability cannot be solely interpreted as current or
voltage-driven. At zero temperature, the (nonuniform)
superconducting state can withstand a current which is
parametrically larger than the thermodynamic depairing current.

High sensitivity of $\Vinst(T)$ to the details of the distribution
function opens avenues for its use as a probe of inelastic
relaxation in the normal state. The shape of $\Vinst(T)$ can be
further used to determine the dominating relaxation mechanism
and extract the corresponding inelastic scattering rate.

We are grateful to M.~V.~Feigel'man, A.~Kamenev, T.~M.~Klapwijk,
J. P. Pekola, V.~V.~Ryazanov,
J.~C.~W.~Song, and D.~Y.~Vodolazov for
discussions. This
work was partially supported by the Russian Federal Agency of Education
(contract No.\ P799) (M. A. S.).

\end{document}